%% file: eprint.tex
\documentclass[10pt, paper=a4, UKenglish]{article}
\usepackage{graphicx}
\def\Title#1{\begin{center} {\Large #1 } \end{center}}
\def\Author#1{\begin{center}{ \sc #1} \end{center}}
\def\Address#1{\begin{center}{ \it #1} \end{center}}

\newcommand\pubblock{\rightline{\begin{tabular}{l} Proceedings of the CTD 2025\\ \pubnumber\\
         \pubdate  \end{tabular}}}

\newenvironment{Abstract}{\begin{quotation} \begin{center} 
             \large ABSTRACT \end{center}\bigskip 
      \begin{center}\begin{large}}{\end{large}\end{center} \end{quotation}}

\newenvironment{Presented}{\begin{quotation} \begin{center} 
             PRESENTED AT\end{center}\bigskip 
      \begin{center}\begin{large}}{\end{large}\end{center} \end{quotation}}

\def\Acknowledgements{\bigskip  \bigskip \begin{center} \begin{large}
      \bf ACKNOWLEDGEMENTS \end{large}\end{center}}

\input econfmacros.tex

\usepackage{color}
\usepackage{lineno}
\usepackage{subfig}
\usepackage{hyperref}
\usepackage{amsmath}

\usepackage[
backend=biber,
style=vancouver,
]{biblatex} 
\newcommand\pubnumber{PROC-CTD2025-071}

\newcommand\pubdate{\today}

\def\affiliation{
Scientific Data Division, Lawrence Berkeley National Laboratory, Berkeley CA, USA\\
\ \\
On behalf of the ATLAS Collaboration}

\newcommand{\conference}{Connecting the Dots Workshop (CTD 2025)\\
November 10-14, 2025}

\usepackage{fancyhdr}
\definecolor{mygrey}{RGB}{105,105,105}
\begin{document}


\large
\begin{titlepage}
\pubblock

\vfill
\Title{Double Metric Learning for Building Directed Graphs
with Chain Connections for the ATLAS ITk Detector}
\vfill

\Author{Jay Chan}
\Address{\affiliation}
\vfill

\begin{Abstract}
Graph construction is an essential step in the Graph Neural Network (GNN) based tracking pipelines. The goal of the graph construction is to construct a graph that contains only the defined true edge connections between nodes (detector hits). A promising approach for the graph construction is through the Metric Learning approach, where a node representation in an embedding space is learned, and nodes are connected according to their distance in the embedding space. The loss function for the metric learning in this case is a contrastive loss encouraging the true pairs of nodes to be close to each other, and pulling away the false pairs of nodes. This approach presents a conflict of the learning objective for the hopping connections when a true edge is defined as a chain connection in a particle track. To address the conflict for this case, we propose a ``Double Metric Learning'' approach, where two node representations are learned. A directed graph can then be constructed based on the distance between the two representations from two nodes respectively. We test this idea with the ATLAS ITk detector at the HL-LHC using the ATLAS ITk simulation and show better graph construction performance particularly for particles with high transverse momentum compared to the Simple Metric Learning approach. We also show that Double Metric Learning is able to accurately predict edge direction.
\end{Abstract}

\vfill

\begin{Presented}
\conference
\end{Presented}
\vfill
\end{titlepage}
\def\thefootnote{\fnsymbol{footnote}}
\setcounter{footnote}{0}
%

\normalsize 


\section{Introduction}
\label{intro}
Graph Neural Networks (GNNs) have emerged as a powerful technique for particle track reconstruction \cite{RevModPhys.82.1419, ATLAS:2017kyn, CMS:2014pgm}, a computationally intensive while critical task in high-energy physics experiments. A typical GNN-based tracking pipeline \cite{ExaTrkX:2021abe, Biscarat:2021dlj, Lieret:2023aqg, ATL-PHYS-PUB-2025-046} begins with a graph construction step designed to build a sparse graph suitable for message passing. In edge-classification frameworks \cite{ExaTrkX:2021abe, Biscarat:2021dlj, ATL-PHYS-PUB-2025-046}, this initial graph also serves as the underlying geometry for subsequent graph segmentation and track formation. Consequently, the fidelity of the constructed graph fundamentally dictates the efficiency and accuracy of the entire downstream reconstruction process.

An effective graph construction algorithm must resolve a tension between two competing objectives: computational tractability and physics performance. Since GNN inference costs scale with edge count, minimizing graph size is essential for latency and memory efficiency. However, the graph must also recover all ``true'' edges to facilitate robust information exchange and ensure successful track segmentation (e.g., via walkthrough algorithms \cite{ATL-PHYS-PUB-2025-046}). Any efficiency loss during construction propagates through the pipeline, manifesting as a direct reduction in final tracking efficiency. This creates a ``tug of war'' between edge purity and signal recall, and the challenge lies in developing algorithms that minimize edge density without sacrificing the underlying physics signal.

Current methodologies for graph construction generally follow either Module Map \cite{Biscarat:2021dlj} or Metric Learning \cite{ExaTrkX:2021abe} paradigms. The Module Map approach connects hits into triplets based on a pre-computed lookup table derived from exhaustive simulation data. To maintain a manageable graph size, geometric cuts are applied to prune these triplets. Conversely, the Metric Learning approach utilizes a neural network to map each hit into a latent embedding space. The model is trained via an objective function that minimizes the distance between hits belonging to a true pair while maximizing the distance between unrelated hits. Edges are then formed by connecting each hit to its neighbors via k-nearest neighbor (k-NN) or fixed-radius nearest neighbor (FRNN) searches. Because the distance metrics in these spaces are typically symmetric, the resulting graphs are inherently undirected.

\begin{figure}[!htb]
  \centering
  \subfloat[]{\includegraphics[scale=0.27]{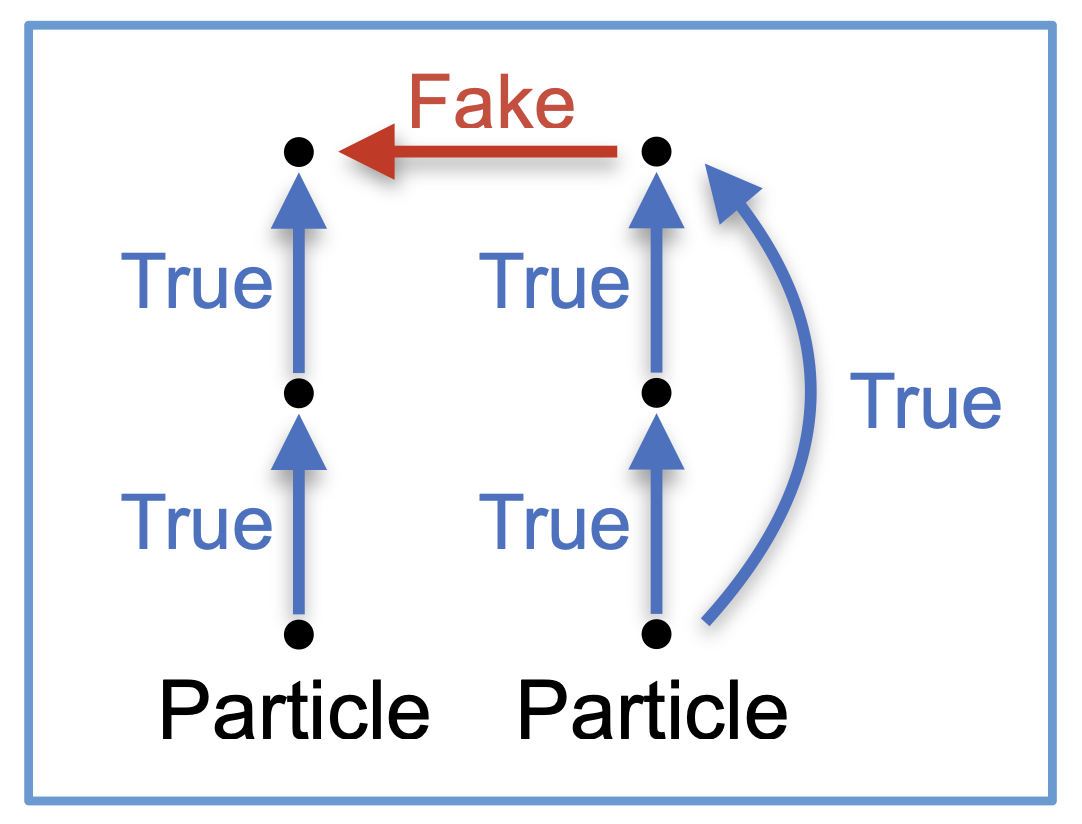}}
  \qquad
  \subfloat[]{\includegraphics[scale=0.27]{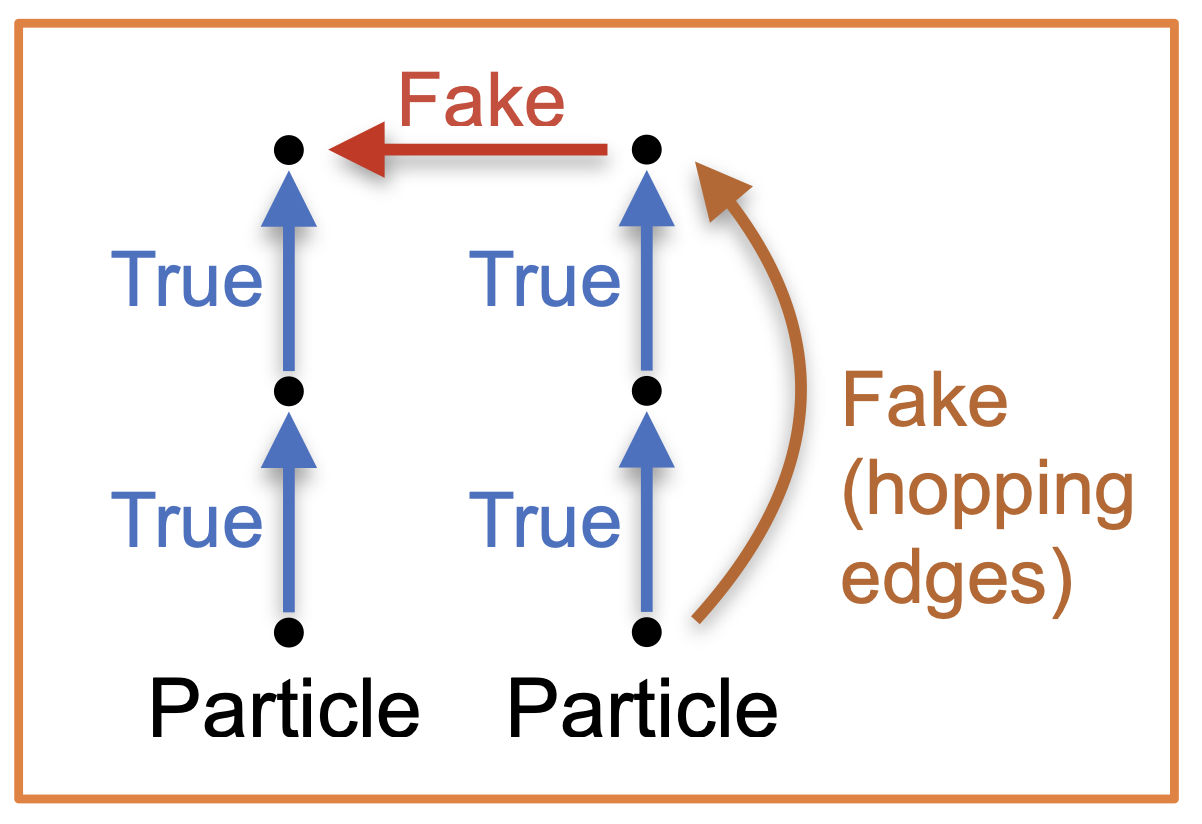}}
  \caption{Illustration of true hit pair definitions: (a) cluster connection, where all hits from the same track are mutually connected; and (b) chain connection, where hits are only connected to their immediate successive neighbors.}
  \label{fig:true_definition}
\end{figure}

The definition of a ``true'' hit pair is central to this learning process and varies by reconstruction strategy. As illustrated in Figure~\ref{fig:true_definition}, under a cluster connection convention, any two hits originating from the same particle are considered a true pair. In contrast, the chain connection definition only recognizes edges between successive hits along a particle’s trajectory. While the choice is application-specific, chain connections are often preferred, as ``hopping'' connections (non-successive hits) are mathematically redundant for message passing and lead to excessive graph density. Furthermore, chain connections are more naturally aligned with the logic of downstream segmentation algorithms like Walkthrough.

In this work, we focus on the Metric Learning paradigm with the chain connection definition and identify a fundamental conflict between standard training objectives and the chain connectivity, which can confuse the training of the machine learning model. To address this, we propose ``Double Metric Learning''---an evolution of the standard approach (hereafter referred to as ``Simple Metric Learning''). A primary advantage of Double Metric Learning is its ability to construct directed graphs, made possible by the inherent directionality of its learned distance metric. We evaluate this method on ATLAS ITk simulation data \cite{ATL-PHYS-PUB-2025-046} and demonstrate that Double Metric Learning significantly enhances graph construction performance, particularly for high-transverse momentum ($p_\mathrm{T}$) particles, while providing high-accuracy edge directionality predictions.

\section{Simple Metric Learning Approach}
\label{sec:SML}

In Simple Metric Learning, a fully-connected neural network is employed to map hit features into a latent representation $\vec{p}$. The neural network is trained to optimize the spatial distribution of these hits in the latent space such that those belonging to a true pair (defined by either cluster or chain connectivity) are clustered together, while unrelated hits are pushed apart. The neural network is trained by minimizing a global loss function $L$, aggregated over a set of representative hit pairs $\{h_i, h_j\}$:
\begin{equation}
    L = \sum_{\{h_i, h_j\}} l\left(h_i, h_j\right).
\label{eq:ml_loss}
\end{equation}
These representative pairs are sampled from three sources:
\begin{enumerate}
    \item True pairs.
    \item Random pairs: Nodes selected randomly to provide a global contrast.
    \item Local neighbors: Each node paired with its neighbors identified via the Fixed Radius Nearest Neighbor (FRNN) algorithm.
\end{enumerate}
The individual loss term $l\left(h_i, h_j\right)$ follows a contrastive hinge loss structure:

\begin{equation}
    l\left(h_i, h_j\right) =
    \begin{cases}
|\vec{p}_i - \vec{p}_j|^2,& \left\{h_i, h_j\right\} \in \text{true pairs} \\
{\max}^2\left(0, m-|\vec{p}_i - \vec{p}_j|^2\right),& \text{otherwise}
\end{cases},
\end{equation}
where $p_i$ and $p_j$ are the latent representations for $h_i$ and $h_j$, respectively, and $m$ is a predefined margin. Minimizing this objective forces the Euclidean distance $|\vec{p}_i - \vec{p}_j|$ toward zero for true pairs, while ensuring that discordant pairs are separated by at least the margin $m$.

Once the latent space is optimized, the graph is constructed by connecting any two hits whose latent distance $|\vec{p}_i - \vec{p}_j|$ falls within a specified radius, determined via FRNN. Since the distance metric is symmetric ($|\vec{p}_i - \vec{p}_j| = |\vec{p}_j - \vec{p}_i|$), the resulting architecture is an undirected graph.

\begin{figure}[!htb]
  \centering
  \subfloat[]{\includegraphics[scale=0.3]{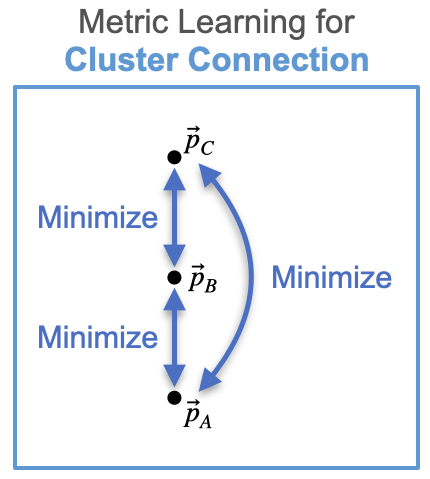}}
  \qquad
  \subfloat[]{\includegraphics[scale=0.3]{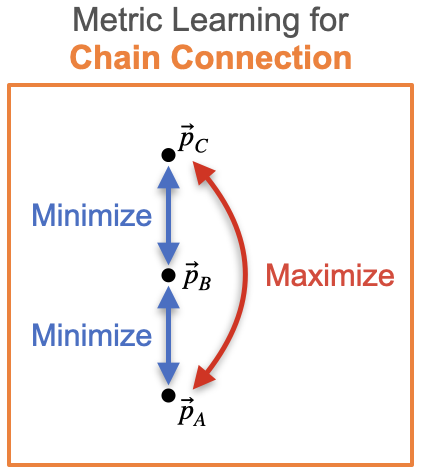}}
  \caption{Schematic of Simple Metric Learning objectives. (a) Cluster connection: forces all track hits into a single cluster. (b) Chain connection: demonstrates the transitive conflict where minimizing adjacent distances ($|\vec{p}_A - \vec{p}_B|$ and $|\vec{p}_B - \vec{p}_C|$) contradicts the maximization of non-adjacent distances ($|\vec{p}_A - \vec{p}_C|$).}
  \label{fig:metric_learning}
\end{figure}

The choice of true pair definition significantly impacts the training dynamics. In the case of a cluster connection convention, the above training objective is ideal for grouping all hits belonging to a single particle track into a single point in latent space. As shown in Figure~\ref{fig:metric_learning}(a), for three hits ($A$, $B$, $C$) on the same track, the network ideally converges to $\vec{p}_A = \vec{p}_B = \vec{p}_C$. On the other hand, the chain connection definition only considers adjacent hits as true pairs. This creates a logical conflict. As illustrated in Figure~\ref{fig:metric_learning}(b), the network attempts to minimize $|\vec{p}_A - \vec{p}_B|$ and $|\vec{p}_B - \vec{p}_C|$, which transitively implies $\vec{p}_A \approx \vec{p}_C$. However, because $\{A, C\}$ is not a true pair in a chain definition, the network simultaneously attempts to maximize $|\vec{p}_A - \vec{p}_C|$ up to the margin $m$.

\section{Double Metric Learning Approach}
\label{sec:DML}

The inherent conflict in Simple Metric Learning with chain connectivity motivates the Double Metric Learning method. In this approach, the neural network instead maps the hit features into the latent space with two distinct representations: a source representation ($\vec{S}$) and a target representation ($\vec{T}$). A hit is projected into its source representation when it acts as the origin of a connection and into the target representation when it acts as the destination. The global loss function is defined over directed pairs ($h_i \to h_j$):

\begin{equation}
    L = \sum_{h_i \to h_j} l\left(h_i \to h_j\right).
\label{eq:ml_loss}
\end{equation}

The individual loss term $l\left(h_i \to h_j\right)$ utilizes the distance between the source representation of the first hit and the target representation of the second:

\begin{equation}
    l\left(h_i \to h_j\right) =
    \begin{cases}
|\vec{S}_i - \vec{T}_j|^2,& h_i \to h_j \in \text{true pairs} \\
{\max}^2\left(0, m-|\vec{S}_i - \vec{T}_j|^2\right),& \text{otherwise}
\end{cases},
\end{equation}
Here, a true pair is strictly defined as a sequence where $h_j$ is the immediate successor of $h_i$ along a particle track. Unlike the simple metric, this relationship is directed. Thus, $h_i \to h_j$ is not equivalent to $h_j \to h_i$.

To construct the graph, a connection is formed from a source hit $h_i$ to a target hit $h_j$ if the latent distance $|\vec{S}_i - \vec{T}_j|$ falls within a specified radius using the FRNN algorithm. Because the distance metric is generally asymmetric under the exchange of hits ($|\vec{S}_i - \vec{T}_j| \neq |\vec{T}_i - \vec{S}_j|$), the resulting architecture is a directed graph. This graph naturally inherits the physical directionality of the particle tracks, aiding in downstream tracking tasks.

\begin{figure}[!htb]
  \centering
  \includegraphics[scale=0.3]{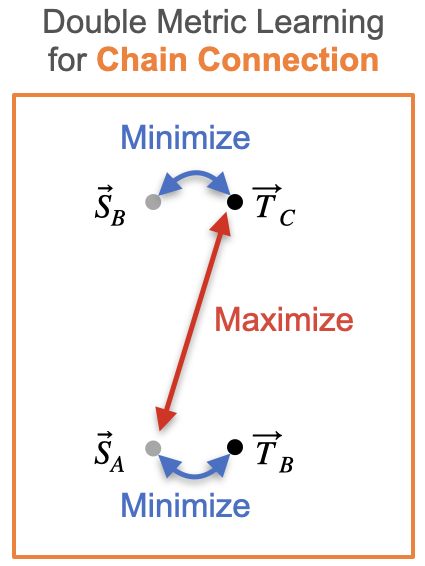}
  \caption{Schematic of the Double Metric Learning objective for chain connection. By utilizing an asymmetric distance metric between dual latent representations, the model satisfies adjacent connectivity without the transitive constraints that cause conflicts in Simple Metric Learning.}
  \label{fig:dml}
\end{figure}

Double Metric Learning successfully resolves the "chain" conflict found in Simple Metric Learning. As illustrated in Figure~\ref{fig:dml}, for a sequence of hits $A \to B \to C$, the network minimizes $|\vec{S}_A - \vec{T}_B|$ and $|\vec{S}_B - \vec{T}_C|$. Crucially, these constraints do not transitively imply that $\vec{S}_A$ must be close to $\vec{T}_C$. Consequently, there is no mathematical contradiction when the loss function simultaneously attempts to maximize $|\vec{S}_A - \vec{T}_C|$ to satisfy the margin requirements for non-successive pairs.

\section{Graph Construction Performance with ATLAS ITk simulation}
\label{sec:result}
To evaluate these methods, we apply both Simple Metric Learning and Double Metric Learning to the ATLAS ITk $t\bar{t}$ simulation (average number of $pp$ interactions per bunch crossing $\langle\mu\rangle = 200$) and compare the structural performance of the resulting graphs. The neural network architectures for both methods are designed to be nearly identical to ensure a controlled comparison. Each model is a multilayer perceptron featuring four hidden layers with a width of 1024 neurons, utilizing tanh activation functions for all intermediate layers.

The primary distinction between the two models lies in the output dimensionality. For Simple Metric Learning, the network produces a single 12-dimensional vector. In contrast, the Double Metric Learning network outputs two separate 12-dimensional vectors representing the source and target representations. In both cases, the final output vectors undergo L2 normalization to project the embeddings onto a unit hypersphere, which stabilizes the distance-based loss. Despite the additional output dimensions in the Double Metric Learning approach, both models contain approximately 3.2 million trainable parameters, resulting in nearly identical computational costs. All models are trained using a learning rate of 0.001 and typically reach convergence within 600 to 700 epochs.

Graphs are constructed using the FRNN algorithm with a radius of 0.12 for Simple Metric Learning and 0.14 for Double Metric Learning. These specific radii are selected to ensure that both methods yield a comparable graph construction efficiency ($\epsilon$), defined as:

\begin{equation}
    \epsilon = \frac{N_\mathrm{Truth}^\mathrm{Reco}}{N_\mathrm{Truth}},
\end{equation}

where $N_\mathrm{Truth}$ represents the total number of ground-truth edges (defined as connections between successive hits) and $N_\mathrm{Truth}^\mathrm{Reco}$ is the number of those true edges successfully reconstructed by the FRNN.

For Simple Metric Learning, the resulting architecture is an undirected graph where edge directions are ignored and redundancies are eliminated. For instance, $A\to B$ and $B\to A$ are treated as a single unique edge. While Double Metric Learning naturally produces a directed graph, we can convert it to an undirected format by ignoring directionality and removing duplicates to facilitate a direct performance comparison between the two methodologies. 

By definition, a directed graph can contain up to twice the number of edges as its corresponding undirected graph, as $A\to B$ and $B\to A$ are treated as distinct. Furthermore, the matching criteria differ: a reconstructed edge in an undirected graph matches a true edge if they connect the same two hits, whereas a directed edge must also match the ground-truth direction. Consequently, the undirected conversion inherently yields a higher construction efficiency than the raw directed graph, as it imposes fewer constraints on the matching process.

\begin{figure}[htb]
    \centering
    \subfloat[][]{\includegraphics[width=0.48\linewidth]{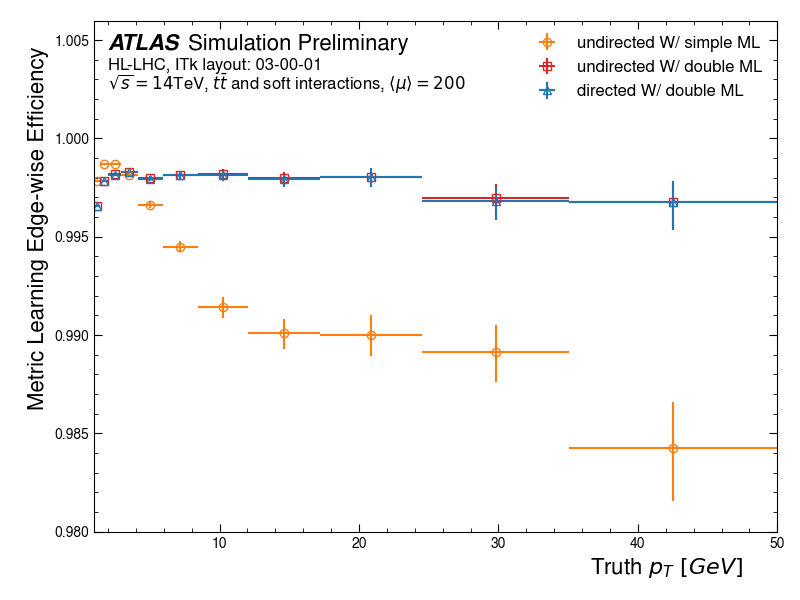}}
    \subfloat[][]{\includegraphics[width=0.48\linewidth]{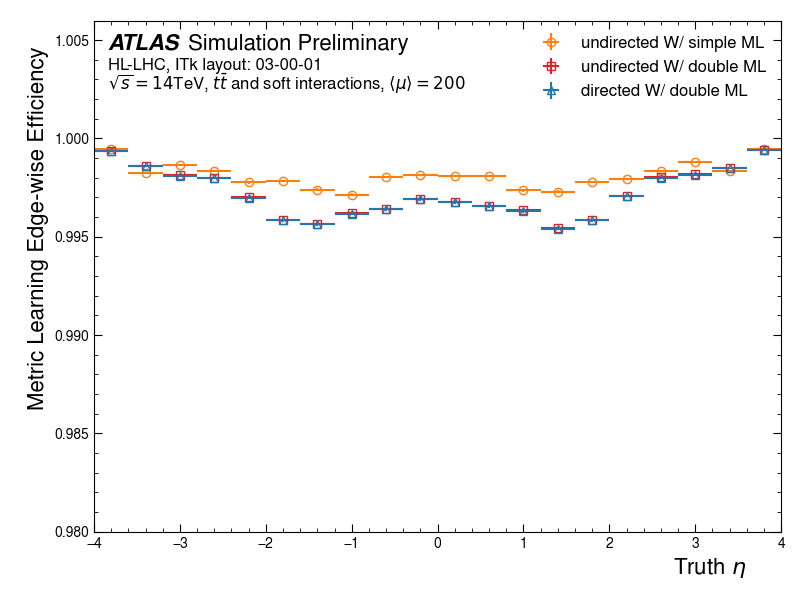}}
    \caption{Graph construction efficiency as a function of (a) the transverse momentum $p_\mathrm{T}$ and (b) the pseudorapidity $\eta$ for hard-scatter particles from $t\bar{t}$ events at $\langle\mu\rangle = 200$, excluding electron tracks, compared between undirected graphs built with Metric Learning (simple ML), undirected graphs built with Double Metric Learning (double ML), and directed graphs built with Double Metric Learning (double ML).}
    \label{fig:dml_eff}
\end{figure}

Table~\ref{tab:constructed_graphs} compares the mean graph size (number of edges) and the total graph construction efficiency of the graphs constructed by both methods. At a comparable level of total construction efficiency, Double Metric Learning produces undirected graphs with a smaller mean graph size compared to Simple Metric Learning. Notably, the directed and undirected versions of the Double Metric Learning graphs exhibit similar edge counts. This indicates that the learned edge directions are highly aligned with the physical directionality of the particle tracks, rarely producing redundant reversed edges. Figure~\ref{fig:dml_eff} shows the graph construction efficiency as a function of the transverse momentum ($p_\mathrm{T}$) and the pseudo-rapidity ($\eta$) of the particles. In particular, Double Metric Learning yields a significantly higher efficiency for particles with higher $p_\mathrm{T}$.

\begin{figure}[htb]
    \centering
    \includegraphics[width=0.5\linewidth]{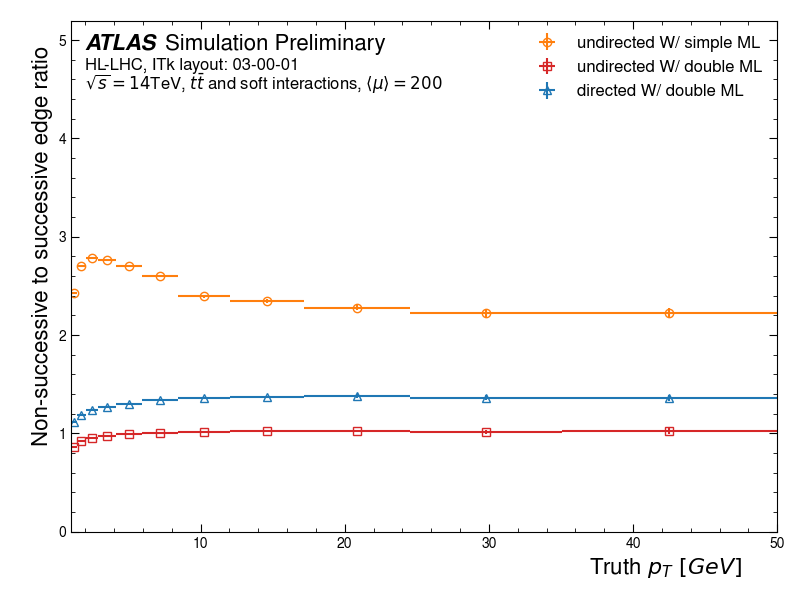}
    \caption{Ratio of number of reconstructed edges corresponding to two non-successive space points (``hopping edges'') from the same particle to the number of reconstructed edges corresponding to two successive space points, as a function of the true transverse momentum $p_\mathrm{T}$.}
    \label{fig:dml_jump_eff}
\end{figure}

To further demonstrate the impact of the Double Metric Learning approach, we examine the ratio of "hopping edges" (reconstructed edges connecting two non-successive hits on a track) to the number of true reconstructed edges in Figure~\ref{fig:dml_jump_eff}. Since only connections between immediate successive hits are defined as true edges, this ratio serves as a metric for the purity of the track-following logic. The results indicate that Double Metric Learning more effectively eliminates these non-physical hopping edges compared to Simple Metric Learning, as the asymmetric distance metric successfully penalizes connections that skip track layers.

\begin{table}[!htb]
  \begin{center}
    \begin{tabular}{l|ccc}
      \hline
      \hline
       &  Mean Graph Size & Total Efficiency ($\epsilon$) \\
      \hline
      Undirected W/ Simple Metric Learning &  $(6.97 \pm 1.52) \times 10^6$  &  0.9981 \\
      Directed W/ Double Metric Learning &  $(7.10 \pm 1.56) \times 10^6$  & 0.9972 \\
      Undirected W/ Double Metric Learning &  $(6.59 \pm 1.45) \times 10^6$  &  0.9972 \\
      \hline
      \hline
    \end{tabular}
    \caption{Comparison of graph construction performance for Simple and Double Metric Learning. The mean graph size (total number of edges) and total graph construction efficiency ($\epsilon$) are shown for the undirected graphs produced by Simple Metric Learning and both the directed and converted undirected graphs produced by Double Metric Learning.}
    \label{tab:constructed_graphs}
  \end{center}
\end{table}

\section{Conclusions}
\label{sec:conclusion}
In this paper, we proposed Double Metric Learning, a novel embedding-based approach for graph construction in particle tracking. By learning dual latent representations, a source and a target representation, for each hit, Double Metric Learning introduces an asymmetric distance metric that effectively resolves the training conflicts inherent in Simple Metric Learning under chain connectivity. Using the ATLAS ITk $t\bar{t}$ simulation, we demonstrated that Double Metric Learning successfully learns edge directions that align with the physical trajectories of particles. Furthermore, the undirected graphs derived from Double Metric Learning exhibit a reduced mean graph size while maintaining high purity, significantly improving construction efficiency, particularly for high-$p_\mathrm{T}$ particles.

Despite these advantages, the current implementation of Double Metric Learning allows for directed graphs that may contain bidirectional edges between the same hits or occasional closed loops due to directionality mispredictions. As certain downstream algorithms, such as walkthrough, require strictly directed acyclic graphs or single-direction edges, future work will focus on developing arbitration logic to resolve these directional ambiguities and enforce a consistent track flow.

Finally, the utility of Double Metric Learning extends beyond edge-classification frameworks. While many current object-condensation frameworks \cite{Lieret:2023aqg} rely on cluster-based connectivity, the transition to Double Metric Learning-based chain connectivity offers a path toward smaller graph sizes. By reducing the number of redundant edges, this approach can lower the computational overhead of message-passing operations and enhance the overall efficiency of graph neural networks in high-luminosity environments.

\Acknowledgements
This work was supported by the DOE HEP Center for Computational Excellence at Lawrence
Berkeley National Laboratory under B\&R KA2401045.

\printbibliography






\end{document}

%% file: econfmacros.tex
\def\beq{\begin{equation}}
\def\eeq#1{\label{#1}\end{equation}}
\def\eeqn{\end{equation}}

\def\beqa{\begin{eqnarray}}
\def\eeqa#1{\label{#1}\end{eqnarray}}
\def\eeqan{\end{eqnarray}}

\let\bar=\overbar

\def\Dslash{\not{\hbox{\kern-4pt $D$}}}
\def\dslash{\not{\hbox{\kern-2pt $\del$}}}

\def\msb{{\bar{\ssstyle M \kern -1pt S}}}

